\magnification1200

\rightline{KCL-MTH-06-14}
\rightline{hep-th/0612001}

\vskip 2cm
\centerline
{\bf Dual fields and  $E_{11}$ }
\vskip 1cm
\centerline{Fabio Riccioni and Peter West}
\centerline{Department of Mathematics}
\centerline{King's College, London WC2R 2LS, UK}
\vskip 2cm
\leftline{\sl Abstract}
We show that the adjoint representation of $E_{11}$ contains generators
corresponding to the infinite possible dual descriptions of the bosonic
on-shell degrees of freedom of eleven dimensional supergravity. We also
give an interpretation for the fields corresponding to many of the  other
generators in the adjoint representation.

\vskip2cm
\noindent

\vskip .5cm

\vfill
\eject
A few years ago it was conjectured [1] that a  rank eleven Kac-Moody
algebra, which was called  $E_{11}$, was a symmetry of M theory.
The non-linear realisation of this algebra at its lowest levels was shown
to contain the    fields of eleven dimensional supergravity and to have
the equations of motion of this theory [1] when one made use of  the
earlier results of reference [2]. The  physical field content is
extracted  by decomposing the adjoint representation of $E_{11}$ into the
representations of its $A_{10}$, or Sl(11), sub-algebra which is
associated in the non-linear realisation with eleven dimensional
gravity.  Non-linear realisations of $E_{11}$ can also be used to describe
ten dimensional theories, but in this case  the adjoint representation is
decomposed into representations of  $A_{9}$, or Sl(10), associated with
ten dimensional gravity. There are only two such $A_9$ subalgebras and
the two different choices were found to  lead to non-linear realisations
which at low levels are the IIA and IIB supergravity theories in
references [1] and [3] respectively.     It is striking to examine tables
of the generators [4]  listed in terms of increasing level  and see how
the generators associated with the field content of the IIA and IIB
supergravity theories occupy precisely all the lower levels before an
infinite sea  of generators whose physical significance was unknown at
the time the tables of reference [4] were constructed.
\par
Amongst the set of
$E_{11}$ generators appropriate to  the IIA theory is one with nine
antisymmetrised indices which in the non-linear realisation leads to a
field with the same index structure [4]. This nine form generator
occurs in the table of IIA generators   at a place which is in amongst
the  generators associated with the fields  of  the  IIA supergravity
theory. A non-trivial value for this  field is known to lead to the
massive IIA supergravity theory and as a result it was realised [4] that
$E_{11}$ can incorporate    the massive IIA  theory.
However, there is a one to one correspondence between the fields of the
non-linear realisations of
$E_{11}$ appropriate to the eleven dimensional, IIA and IIB theories  and
using this correspondence one finds  that the nine form of the IIA theory
corresponds to a field with the index structure
$A^{a,b,c_1\ldots c_{10}}$ which occurs at a level which is beyond those
of the supergravity fields of eleven dimensional supergravity theory
[5].   As such $E_{11}$ provides for an eleven dimensional origin of the
massive IIA theory which involves one of the higher level fields and in
this way the physical interpretation of at least one of the higher level
fields in the eleven dimensional $E_{11}$ theory became apparent.
\par
The non-linear realisation of $E_{11}$ appropriate to the IIB theory
was observed to contain a number of eight form and ten form fields
at levels just  above  the usual fields of supergravity [4]. The IIB
supergravity theory was originally constructed by closing the local
supersymmetry algebra [6].  One can add  to the standard supergravity
fields  a SL(2,R) doublet of six forms, dual to the two forms, and a
triplet of eight forms dual to the scalars, whose 9-form field strengths
satisfy a constraint so that only two fields actually propagate [9],
but in an unexpected recent development [7] it was shown that one could
also add a very specific set of ten forms to the theory and still close
the supersymmetry algebra. Remarkably, these
fields were in precise correspondence with those already found in
$E_{11}$ [4]. Furthermore,   the exact  coefficients of their   gauge
transformations deduced from supersymmetry closure [7] were found [8] to
agree with their derivation from the $E_{11}$ algebra. The  ten
form fields, although non-propagating,  have an important role
since they are associated with spacetime-filling branes. In particular,
one of the 10-forms in the IIB algebra is associated to D9-branes, that
take part in the orientifold projection  which gives rise to the
type-I theory from the IIB theory [10]. Thus a few more
of the higher level fields were found to have a physical meaning. The
non-linear realisation of
$E_{11}$ appropriate to IIA theory was also observed  to contain some ten
forms [4]  and these were recently found [11] to be in precise agreement
with the ten forms one could add and still have a closing supersymmetry
algebra in the IIA theory.
\par
In this paper we will  consider  the generators of the adjoint
representation of $E_{11}$ classified according to the $A_{10}$
sub-algebra corresponding to an   eleven dimensional gravity subsector.
We  will find all generators that have no blocks of $A_{10}$ indices
containing ten antisymmetrised indices and whose indices sum to $3l$
where $l$ is the level of the generator. The lowest level such generators
have the index structure  $ R^{abc}$, $R^{a_1\ldots a_6}$
$ R^{a_1\ldots a_8,b}$ and  correspond to the three and six form gauge
fields of the supergravity theory and the dual graviton field.
Although there could in principle be  a very large
number of  generators in the class being considered we will show  that it
only  contains  the above generators and the above generators
decorated by blocks each of which contain nine  indices.
For example, for the three form this means the index structure
$ R^{abc, d_1\ldots d_9, e_1\ldots e_9 }$ when we have only two blocks
of nine indices.
The fields corresponding to these  generators are
just all possible dual formulations of the bosonic  on-shell degrees of
freedom of eleven dimensional supergravity. We interpret this  result to
mean that this sector of the adjoint representation of $E_{11}$  encodes
the most general possible duality symmetries of bosonic eleven dimensional
supergravity in a manifest manner.  As a result, we will find a physical
interpretation for an infinite number of $E_{11}$ generators.
\par
Very little is known about generalised  Kac-Moody algebras, indeed, for
no such algebra is there even a listing of the generators that are present
in the algebra. However,   Kac-Moody algebras   which  possess  at least
one node whose deletion leads to a set of Dynkin diagrams each of which
corresponds  to one of the finite dimensional semi-simple Lie algebras
are more amenable to analysis [12]. In particular,  one can analyse the
Kac-Moody  algebra in terms of the finite dimensional sub-algebras that
appears after this deletion. The Dynkin diagram of
$E_{11}$ is given by
$$
\matrix{ & & & & & & & & & & & & & &0&11&\cr
        & & & & & & & & & & & & & &|& &\cr
       0&-&0&-&0&-&0&-&0&-&0&-&0&-&0&-&0&-&0\cr
       1& &2& &3& &4& & 5& &6& &7& &8& &9& &10\cr}
$$
In the case of
$E_{11}$, if one deletes  the node labeled eleven then the remaining
algebra is
$A_{10}$ or SL(11). We call this $A_{10}$ section of the Dynkin diagram
the gravity line as it gives rise  in the non-linear realisation to eleven
dimensional gravity sector of the theory.  It will prove useful to
recall how one analyses the generators of the adjoint representation
$E_{11}$ in terms of representations of $A_{10}$ [12,13,14].  The simple
roots of
$E_{11}$ are the  simple
roots  $\alpha_i,\ i=1,\ldots , 10$ of $A_{10}$ as well as the simple
root corresponding to node eleven which we call $\alpha_c$ which is given
by
$$
\alpha_c=x-\lambda_8
\eqno(1)$$
Here $x$ is orthogonal to the roots $\alpha_i,\ i=1,\ldots , 10$ of
$A_{10}$ and  $\lambda_i, ,\ i=1,\ldots , 10$ are the fundamental weights
of  $A_{10}$
As   $\alpha_c\cdot \alpha_c=2$ and $\lambda_8\cdot \lambda_8={8.3\over
11}$,  we find that
$x^2=-{2\over 11}$. Any root
$\alpha$  of
$E_{11}$ can be expressed in terms of its  simple roots of
$E_{11}$ and, using equation (1), can be written as follows
$$
\alpha= \sum_{i=1}^{10} n_i\alpha_i +l\alpha_c= lx- \Lambda
\eqno(2)$$
where
$$
\Lambda = l\lambda_8- \sum_{i=1}^{10} n_i\alpha_i
\eqno(3)$$
is in the weight space of $A_{10}$.
\par
We call $l$ the level of the root and the strategy [12,13,14] is to
analyse the generators in the adjoint representation of $E_{11}$ level by
level in terms of the representations of $A_{10}$. When
constructing the $E_{11}$ algebra in terms of multiple commutators using
the Serre relation,  the level is just the number of times the generator
$R^{abc}$ occurs in the commutators.
\par
If the adjoint representation of
$E_{11}$ contains the representation of
$A_{10}$ at level $l$ with Dynkin indices $p_j$, then we must find that
some root of $E_{11}$ contains an $A_{10}$ weight  of the form
$\Lambda=\sum_j p_j\lambda_j$ where $p_j$ are  positive integers. As such,
one finds that
$$
\sum_j p_j\lambda_j= l\lambda_8-\sum_j n_j\alpha_j
\eqno(4)$$
Taking the scalar product with $\lambda_i$, and using the equation
$$(A_{ik})^{-1}=(\lambda_i,\lambda_k)
\eqno(5)$$
valid for any simply laced finite dimensional semi-simple Lie algebra,
implies that
$$
\sum_j p_j A_{ji}^{-1}= l A_{8i}^{-1}- n_i
\eqno(6)$$
where  $A_{ji}^{-1}$ is the inverse Cartan matrix of $A_{10}$. After
some though one can see that this must also occur for
$p_j$, $l$ and $n_i$ are all positive integers. As such equation (6)
places strong restrictions on the possible $p_j$, and so the
representations of $A_{10}$,  that can occur at a given level. This
observations relies on the fact that the inverse Cartan matrix is positive
definite for any finite dimensional semi-simple Lie algebra.
\par
Any Kac-Moody algebra with
symmetric Cartan matrix has its roots bounded by $\alpha^2 \le
2,0, -2\ldots $ [15]. Consequently, using equations
(2) and (5) we find that
$$\alpha^2=l^2 x^2+\sum_{ij}p_i (A^{}_{ij})^{-1} p_j\le 2,0,-2,\ldots
\eqno(7)$$
Since the first term is fixed for a given $l$ and the second term is
positive definite this also places constraints on  the possible values of
$p_i$.
\par
It will be useful to  recall the generators of $E_{11}$ that occur
for the first three levels [1]. At level zero we have the generators
$K^a{}_b$ of
$A_{10}$, at level one
$$ R^{abc}, \ \ l=1,\ p_8=1\ \ \alpha_A=(0,0,0,0,0,0,0,0,0,0,1)
\eqno(8)$$
at level two
$$R^{a_1\ldots a_6},\ \  l=2,\ p_5=1\ \ \alpha_B=(0,0,0,0,0,1,2,3,2,1,2)
\eqno(9)$$
at level three
$$ R^{a_1\ldots a_8,b},\ \ l=3,\ p_{10}=1,p_3=1\ \
\alpha_C=(0,0,0,1,2,3,4,5,3,1,3)
\eqno(10)$$
The vectors shown above are the  $E_{11}$ roots, the entries being the
integers of equation (2). At level three  we also find
$$
R^{a_1\ldots a_9},\ \ p_2=1\ \ \alpha_D=(0,0,1,2,3,4,5,6,4,2,3) ,
\eqno(11)$$
but this generator has multiplicity zero and so does not actually
occur in the $E_{11}$ algebra.
All other $p_j$'s not mentioned  in the above expressions are zero.
In deriving these results one uses that the inverse
Cartan matrix of
$A_{n-1}$, or SL(n),  is given by
$$(A_{jk})^{-1}=\cases{{j(n-k)\over n}, \ \ j\le k\cr
{k(n-j)\over n}, \ \ j\ge k\cr}
\eqno(12)$$
\par
We now show that the $E_{11}$ generators at  level $l$ when decomposed
into  representations of $A_{10}$ can be written in such a way that they
always have
$3l$ indices. This includes level zero if one counts upper indices with a
plus sign and lower indices with a minus sign. Clearly, this counting is
not respected when changing indices using the epsilon symbol. Like all
Kac-Moody algebras,
$E_{11}$  is just the multiple commutators of   its Chevalley generators,
subject to the Serre relation. The multiple commutators of the  level zero
Chevalley generators lead to  SL(11), or
$A_{10}$ that is the generators $K^a{}_b$. While   the multiple
commutators of these,  together with one of the one level one
Chevalley generator, $R^{91011}$  lead to the generators $R^{abc}$. The
rest of the positive root part of the algebra is then found from multiple
commutators of
$R^{abc}$. As a result,  if one never uses epsilons on the right hand
side of the commutators one finds generators that have
$3l$ SL(11) indices.    The same applies to the construction of all the
negative root generators which can also be written with  $3l$ indices,
but now the levels are negative. This is  consistent with the assignment
of a lower index with $-1$ contribution.
\par
The generators that emerge from the above analysis are representations of
$A_{10}$ and so are labeled by their associated Dynkin indices $p_j$.
In our conventions a generator with $p_j=1$ has $11-j$ antisymmetrised
indices. To each non-vanishing Dynkin index, $p_j$ the generator possess
$p_j$  blocks of antisymmetric indices. The blocks of antisymmetrised
indices are in general  subject to  symmetrisation relations.
\par
We now divide all the generators  into two classes,  those that have
at least one block of indices that contain  ten or eleven antisymmetrised
indices and those generators that do not contain any such blocks. We refer
to the latter generators as {\bf dual} generators for a reason that will
become apparent.   We note that a block of eleven antisymmetrised indices
is equivalent to a scalar and so does not correspond to   any Dynkin
index.  However, one can count the number of such blocks as it is just
determined  by the mismatch between  the sum of the indices that arise
from the blocks associated with the Dynkin indices $p_j$  and  $3l$. The
dual  generators have, by definition, no such rank eleven index blocks
and also have the Dynkin index
$p_1=0$, as they also  have no blocks of indices with ten antisymmetrised
indices. As such, dual generators have  $p_1=0$ and sum of all the
indices must be
$3l$;
$$
\sum_{j=2}^{10} p_j(11-j)=3l
\eqno(13)$$
Let us give some more  examples. At level four in the table of
$E_{11}$ generators we find a generators with Dynkin indices $p_2=1$ and
$p_8=1$, all other Dynkin indices being zero. This has the index
structure
$R^{a_1a_2a_3, b_1\ldots b_9}$,  it has the required $3.4=12$ indices and
so is a dual generator. Also at level four we find a generator with
$p_{10}=1$, all other Dynkin indices being zero. This must be of the
form  $R^{a_1, b_1\ldots b_{11}}$ in order to have the required $12$
indices and so it is not a dual generator.
\par
We now use  equations (6) and (7) to see
what constraints  they place on the possible dual generators. We begin
with the condition on the length of the root of equation (7). Substituting
for $l$ from equation (13) we find that  equation (7) considerably
simplifies and can be written as
$$
\alpha^2= {1\over 9} \sum _j (11-j)(j-2) p_j^2
+{2\over 9}\sum_{i<j}(11-j)(i-2) p_i p_j=2,0,-2,\ldots
\eqno(14)$$
We observe that the left hand side is positive definite and so
the root $\alpha$ can only have length squared $2$ or $0$. Furthermore, we
note that the equation does not contain  $p_2$ and so places no
constraint on it. It is straightforward to show that the only solutions
are
$$
{\rm solution\  A},\ \ \ \ \ p_2=u, p_8=1,\  \ \ {\rm at} \ \ l=1+3u,
\eqno(15)$$
$$
{\rm solution\  B},\ \ \ \ \ p_2=u, p_5=1, \  \ \ {\rm at}\ \ l=2+3u,
\eqno(16)$$
$$
{\rm solution\  C},\ \ \ \ \ p_2=u, p_3=1=p_{10},\  \ \ {\rm  at
}\ \ l=3+3u,
\eqno(17)$$
$$
{\rm solution\  D},\ \ \ \ \ p_2=u+1,,\  \ \ {\rm  at }\ \ l=3+3u,
\eqno(18)$$
\par
All these solutions automatically satisfy equation (6) and one finds that
the $E_{11}$ roots they correspond to are given by
$$
\alpha _I(u)=\alpha_I+u\alpha_D
\eqno(19)$$
where the label $I=A, B, C, D$ corresponding to the solutions
and $\alpha_I$  are given in equations (8-11).
\par
We first discuss the solutions A,B,C which all have $\alpha^2=2$. These
contain just  one solution  at every level. The solutions for $u=0$ are
just the three
form $R^{abc}$ at level one, the six form  $R^{a_1\ldots a_6}$ at level
two and the dual graviton generator  $R^{a_1\ldots a_8,b}$ at level three
given in equations (8), (9) and (10). This must be the
case as all these generators are dual generators.
At  higher levels this pattern is repeated, but one finds instead
that the generators acquire  $u$ blocks of nine  indices.
Solution D had $\alpha^2=0$ and for  $u=0$ it is just the
nine  form at level three of equation (11).  At all higher levels spaced
by three one finds in  solution $D$   generators with $u+1$ blocks
of nine indices. In fact the  level three, i.e. $u=0$,    nine  form
generator $R^{a_1\ldots a_9}$, has multiplicity zero and examining the
$E_{11}$ tables it would appear that all the other generators  in solution
$D$ also have multiplicity zero  up to level nine and one  can expect
this to be true at  all the levels. This means that although these
generators solve all the constraints of equations (6) and (7) they  do
not actually occur in the algebra. The constraints of equations (6) and
(7) are a necessary,  but not a sufficient condition. What generators
actually occur in the algebra  are
determined by multiple commutators of the Chevalley relations subject to
the Serre relation and this is more restrictive that the above
conditions. In contrast, looking at the tables of reference [16,19]   we
see that  the generators for solutions A, B and C all have multiplicity
one up to level ten and so do occur in the algebra. We can expect that
they have multiplicity one at all levels.
\par
Although equations (6) and (7) can be used to  systematically
search for a generator with fixed Dynkin indices at all
levels, there  is usually little hope of successfully carrying out such a
calculation unless the Dynkin indices are specified.  However, this was
not the case for the dual generators indicating that they are preferred
objects in the
$E_{11}$ algebra.
\par
The massless states of the eleven dimensional Poincare
algebra are classified in terms of representations of its little group
SO(9).   If we take the indices on the generators in the solutions A, B
and C in equations (15-17) to only take the values
$1,\ldots , 9$,   the blocks of nine  indices  are equivalent to no
indices at all  and so we find only the three rank, six rank and the
$R^{j_i\ldots j_8, i}$ irreducible representations of SO(9).  Of course
the rank three and rank six generators are the same representations of
SO(9) as is apparent by invariant epsilon symbol. The same is true for
the graviton contained in the level zero sector and the
$R^{j_i\ldots j_8, i}$ generator at level three as the latter
is just equivalent to the generator
$R_j{}^i$. In fact the generator
$R^{j_i\ldots j_8,i}$ obeys the constraint
$R^{[j_1\ldots j_8, j_9]}=0$ and so the equivalent generator obeys
$R_i{}^i=0$. The fields corresponding to
the generators of the solutions A, B and C are just all possible dual
descriptions of the  bosonic  degrees of freedom of the eleven
dimensional supergravity theory. The meaning  of this result is that the
sector of the $E_{11}$ non-linear realisation associated  with the
dual generators encodes all possible duality symmetries in a manifest
way. We note that all the non-dual fields vanish when we restrict their
indices to only take the values $1,\ldots , 9$ and so they lead
to no representations of the little group SO(9). This suggests that the
non-dual generators  lead to no propagating degrees of freedom.
\par
The roots of equation (19) for the generators in the solutions of
equations  (15-17) are consistent with the conjecture that
all these generators are found taking multiple commutators of  the
generator $R^{a_1\ldots a_9}$ of equation (18) with the
three generators of equa\-tions
(8-10).
\par
We now apply  similar considerations to the $E_{11}$  theory
appropriate to the IIB  theory.  The Dynkin diagram of $E_{11}$
with  the $A_9$ subgroup associated with the gravity sector of the IIB
theory shown as the horizontal line (the so called gravity line) [3].
$$
\matrix{ & & & & & & & & &  & & & &  &0&10 &\cr
         & & & & & & & & & & & & & &|& &\cr
         & & & & & & & & & & & & &  &0&9 &\cr
          & & & & & & & & & & & & & &|& &\cr
        0&-&0&-&0&-&0&-&0&-&0&-&0&-&0&-&0\cr
        1& &2& &3& &4& & 5& &6& &7& &8& &11\cr}
$$
It is immediately apparent that the IIB theory has an SL(2,R)
invariance which arises from the node labeled ten as this is not
connected to the gravity line and so corresponds to an internal
symmetry. For this theory we delete the node labeled 9 and then the
algebra splits into
$A_9\otimes A_1$. The simple roots of
$E_{11}$ are the simple $\alpha_i,\ i=1,\ldots , 8,11$ of $A_9$, the root
$\alpha$ of $A_1$ and the root  corresponding to node 9 which we call
$\alpha_c$ which  is given by
$$
\alpha_c=x-\mu-\lambda_8
\eqno(20)$$
where $\lambda_i,\ i=1,\ldots , 8,11$ are the fundamental weights of $A_9$
and $\mu$ is the fundamental weight of $A_1$. We note that the last node
in the gravity line is node 11 and we keep this labeling for the $A_{9}$
Dynkin diagram.  As
$\alpha_c\cdot\alpha_c=2$ we find that
$x^2=-{1\over 10}$. Any root
$\alpha$  of $E_{11}$ can be expressed in terms of the simple roots of
$E_{11}$  as follows
$$
\alpha= \sum_{i=1}^{8,11} n_i\alpha_i +l\alpha_c+m\alpha= lx-
\Lambda^{{A_9}} -\Lambda^{{A_1}}
\eqno(21)$$
where
$$
\Lambda^{{A_9}} = l\lambda_8- \sum_{i=1}^{8,11} n_i\alpha_i,\ \ {\rm and
}\
\ \Lambda^{{A_1}} = l\mu- m\alpha
\eqno(22)$$
We note that  $\Lambda^{{A_9}}$ and $\Lambda^{{A_1}}$ are in the weight
spaces of
$A_9$ and $A_1$ respectively. A necessary condition for  the adjoint
representation of
$E_{11}$ to contain the representation of
$A_9\otimes A_1$,  with Dynkin indices $p_j$ and $q$ for each factor
respectively, is that
$$\Lambda^{{A_9}} =l\lambda_8- \sum_{i=1}^{8,11}
n_i\alpha_i=\sum_jp_j\lambda_j,
\eqno(23)$$
and
$$
\Lambda^{{A_1}} = l\mu- m\alpha=q\mu
\eqno(24)$$
Where $l,n_i,p_i$ and $m,q$ are all positive integers or zero. Taking the
scalar product with the fundamental weights,  these two equations
respectively become
$$
\sum_j p_j A_{ji}^{-1}= l A_{8i}^{-1}- n_i
\eqno(25)$$
and
$$
l-q=2m
\eqno(26)$$
where now $A_{ji}^{-1}$ is the inverse Cartan matrix of $A_9$. The
IIB decomposition given in reference [4] has two levels $l_1$ and $l_2$
corresponding to the deletion of the nodes ten  and nine. The
decomposition studied here has the advantage that it automatically works
with complete representations of SL(2,R). For comparison  we have  $l_2=l$
while $l_1=m={1\over 2}(l-q)$
\par
The condition on the length of the roots is given by
$$
\alpha^2=l^2 x^2+\sum_{ij}p_i (A_{ij})^{-1} p_j+{q^2\over 2}\le
2,0,-2,\ldots
\eqno(27)$$
\par
We now divide the generators of the IIB theory into two classes in a
way that is similar to that done above   for the eleven dimensional
theory. The generators of the  IIB theory can all be written in
such a way that they carry $2l$ indices. This is essentially because the
Chevalley generators of the IIB theory are those  associated
with the gravity line nodes one to eight  and node eleven,  whose
multiple commutators lead to the generators  $K^a{}_b$ of $A_9$, the
Chevalley generator of node ten which is associated with SL(2,R),  whose
multiple commutators introduce no space-time indices and finally   the
Chevalley generator of node nine which is contained in one of the two
generators of the form $R^{a b}$ and has level one. It is the multiple
commutators of this latter generators that lead to  the addition of two
space-time indices at every level.  We now call dual generators
those that have no blocks of antisymmetrised indices of rank nine or
ten.  As a result,  the dual generators          have
$p_1=0$ and
$$
\sum_{j=2}^{8,11} p_j(10-j)=2l
\eqno(28)$$
Using equation (28) we find the length condition on the root of equation
(27) leads to the constraint
$$
\alpha^2= {1\over 8} \sum _j (10-j)(j-2) p_j^2
+{1\over 4}\sum_{i<j}(10-j)(i-2) p_i p_j+{q^2\over 2}=2,0,-2,\ldots
\eqno(29)$$
We see that the left hand side is positive definite and so the only
non-trivial solutions have $\alpha^2=2$. We also observe that the Dynkin
label $p_2$ does not appear and so it not subject to any constraint.
Furthermore $q$ can only take the values $q=0,1,2$. If we have $q=2$ then
all the $p_j=0$ except $p_2$ which can take any value.
It is easy to see that the only solutions, which also satisfy equations
(25) and (26),  are
$$
{\rm solution\  A},\ \ R^{a_1a_2},  \ \ \ p_2=u, p_8=1,\  q=1,\  \
\  \ {\rm at} \ \
l=1+4u,
\eqno(30)$$
$$
{\rm solution\  B},\ \ R^{a_1\ldots a_4},  \ \ \ p_2=u, p_6=1,\ q=0 \  \
\  \ {\rm at}\ \ l=2+4u,
\eqno(31)$$
$$
{\rm solution\  C},\ \ R^{a_1\ldots a_6}, \ \ \ p_2=u, p_4=1,\ q=1\  \
\ \ {\rm  at}\ \ l=3+4u,
\eqno(32)$$
$$
{\rm solution\  D},\ \ R^{a_1\ldots a_7,b},\ \ \ p_2=u, p_3=1=p_9,\ q=0\
\ \ \ {\rm  at}\ \ l=4+4u,
\eqno(33)$$
$$
{\rm solution\  E},\ \ R^{a_1\ldots a_8},\ \ \ p_2=u+1 ,\ q=2\  \
\ \ {\rm
at}\ \ l=4+4u,
\eqno(34)$$
$$
{\rm solution\  F},\ \ R^{a_1\ldots a_8},\ \ \ p_2=u+1 ,\ q=0\  \
\ \ {\rm
at}\ \ l=4+4u,
\eqno(35)$$
The above only lists the highest weight components of each SL(2,R)
multiplet and one can obtain all members of the multiplet by acting with
the Chevalley generators $E_{10}$ and $F_{10}$. The axion is not
displayed as it is part of the SL(2,R) multiplet which contains the
dilaton as its highest weight component. We notice that
for $u=0$ we have precisely the generators corresponding to the content of
the IIB supergravity theory if we include the dual graviton and the eight
form generators. The above solutions  all have $\alpha^2=2$ except
for solution F which has $\alpha^2=0$. The
general solution above has generators which have the indices of the $u=0$
generators    decorated by any number of blocks of eight indices.
The roots are found from equations (25) and (26). The simplest way to
deduce them is to first show that if one has a solution with certain set
of Dynkin indices $p_j$, then there exists another solution with the same
Dynkin indices except with $p_2$ replaced by $p_2+u$ for $u$ a positive
integer.  The resulting $E_{11}$ root $\alpha$ then changes   to
$\alpha+u\alpha_F$.  Hence,  given the above solutions for $u=0$ we
generate all the others  and they have the roots
$$
\alpha_I(u)=\alpha_I+u\alpha_F
\eqno(36)$$
where $I$ stands for A,B,C,D,E and F and the roots $\alpha_I$ are given
by
$$
\alpha_A=(0,0,0,0,0,0,0,0,1,0,0),\ \ \alpha_B=(0,0,0,0,0,0,1,2,2,1,1),
$$
$$
\alpha_C=(0,0,0,0,1,2,3,4,3,1,2),\ \ \alpha_D=(0,0,0,1,2,3,4,5,4,2,2),
$$
$$
\alpha_E=(0,0,1,2,3,4,5,6,4,1,3),\ \  \alpha_F=(0,0,1,2,3,4,5,6,4,2,3)
\eqno(37)$$
\par
The little group in ten dimensions is SO(8). The above solutions have a
similar interpretations as the ones in eleven dimensions. They correspond
to all possible dual descriptions of the  on-shell bosonic degrees of
freedom of the IIB supergravity theory and so one expects that the
non-linear realisation corresponding to the IIB theory manifestly
displays all possible duality symmetries corresponding to the on-shell
degrees of IIB super\-gravity. As for the eleven dimensional case, the
non-dual fields do not contribute any non-trivial little
group representations.   This completes our discussion of the IIB case.
\par
\par
We observed that for the eleven dimensional
case the dual generators were likely to be result from
taking repeated commutators of the generator at level three which had
$p_2=1, l=3$,  $\alpha^2=0$ and multiplicity zero with
the three lowest level generators.
In fact such a result applies not only to the dual generators, but to all
the $E_{11}$ generators as we now explain.  Given two  solutions with
Dynkin indices and levels given by
$(p^{(1)}, l^{(1)})$ and
$(p^{(2)}, l^{(2)})$  to the weight matching condition of equation (6),
then we find that  $(p^{(1)}+(p^{(2)}, l^{(1)} + l^{(2)})$  is also a
solution as this equation is linear in its Dynkin indices and level.
This is not in general true  for the   root length
condition of equation (7). However, if we take   $(p^{(2)},
l^{(2)})$ to be the generator of equation (18) with $u=0$, i.e.
$(p^{(2)}_2=1, l^{(2)}=3)$, then we do find a solution of equation (7) of
the form
$(p^{(1)}+p^{(2)}, l^{(1)} + l^{(2)})$. Indeed,
if we denote the root of  the latter   by $\beta$ then
it has a root length squared which  is  given by
$$
(\beta)^2=\alpha_1^2-{2.2\over 11}l^{(1)}l^{(2)}
+\sum_{i,j} p_j^{(1)} A_{ji}^{-1} p_j^{(2)}
=\alpha_1^2-2n^{(1)}_2
\eqno(38)$$
which is always in the allowed range. In going between the two equations
we have used equation (6) for the first solution. Thus even though the
generator with  $p_2=1, l=3$ does not actually appear in the $E_{11}$
algebra its repeated  commutator with any generator in the
$E_{11}$ algebra should result in   another generator in the algebra.
Examining the table of $E_{11}$ generators [16,19] one does indeed  find
generators which are genuine members of the $E_{11}$ algebra with the
correct structure  to have arisen from such commutators.
Such generators will have the same SO(9) little group representations as
the original generator before any commutators and as such it would seem
that all fields in the non-linear realisation of $E_{11}$ come in all
possible dual formulations.
\par
The $E_{11}$ root corresponding to  the generator $R^{a_1\ldots a_9}$
with
$p_2=1, l=3$ is of the form $k\equiv\alpha_3+\theta$ where $\theta $ is
the highest root of $E_8$ and 
$\alpha_3$ is the root whose addition  corresponds to changing  $E_8$
to the affine algebra $E_9$. As such,  $k$   is associated with the null
root of the affine $E_9$ subalgebra. We note that in any affine algebra
the null root
$k$ is responsible for the infinite number of generators in the algebra
as if $\alpha$ is a root of length squared two found in the underlying
finite dimensional algebra then the roots of the algebra are of the form
$\alpha+nk$. The infinite number of generators in the affine algebra can
be packaged up into
$T^a(z)=T^a_n z^n$ where $z$ parameterises the $S^1$ of the circle or
loop.  Given the above results we conclude that we can package all the
generators of $E_{11}$ in a similar manner by considering them as 
coefficients of a polynomial in the variable $z^{a_1\ldots a_{9}}$. 
Even once this packaging has been carried out the adjoint
representation contains an infinite number of  generators even though they
are functions of $z^{a_1\ldots a_{9}}$. However, the dual generators can
all be encoded by taking the generators of the adjoint representation of
$E_8$ and letting them depend on $z^{a_1\ldots a_{9}}$.   Could it be
that $E_{11}$ can be formulated as a mapping from a three dimensional
surface into the $E_8$, algebra and could the same  apply to any
very extended algebra $G^{+++}$?   
\par
Similar results to those discussed just above hold  for the IIB theory
using  the generator of equation (35).
\par
Given any finite dimensional semi-simple Lie algebra $G$ one can form a
unique rank three larger Kac-Moody algebra $G^{+++}$ based upon it [11]
and consider the non-linear realisation formed from it [1,17,18]. For
example, if $G=E_8$ then $E_8^{+++}=E_{11}$. The low
level dynamics of these non-linear realisations are many of the theories
of interest to theoretical physicists. For example,  the non-linear
realisations of
$A_{D-3}^{+++}$  and
$D_{D-2}^{+++}$  are gravity  [1] and a generalisation of the
effective action of the  bosonic string  [17] in $D$ space-time dimensions
respectively. Examining the tables of references [4] and [19] one finds
that the story of the dual generators found in the context of $E_{11}$
also holds for these theories. Namely, the dual generators are
the original low level generators associated with the physical
fields of the theory as well as these generators when suplimented by
blocks of
$D-2$ indices. Furthermore there exists in each case   generators of
length squared zero whose multiple commutators with the very lowest dual
generators are likely to lead to all dual generators. This is particularly
apparent when looking at the table of generators of $D_8^{+++}$ and
$B_3^{+++}$ of reference [4] and $A_1^{+++}$ of reference [19].
\par
Let us summarise the results in this paper,  we have considered the
adjoint representa\-tion of $E_{11}$ when decomposed with respect to the
$A_{10}$ sub-algebra corresponding to the gravity sector of  an  eleven
dimensional theory. We have explained that all the generators of $E_{11}$
can be written with
$3l$ indices where $l$ is the level of the generator. We then found
all generators of $E_{11}$ that do not have any blocks of ten or eleven
antisymmetrised indices.  There is
only one such generator at every level and they correspond to all
possible dual formulations of the on-shell bosonic degrees of freedom of
the eleven dimensional supergravity theory.  Thus we have found a
physical meaning for an infinite number of $E_{11}$ generators.
\par
The
non-dual generators, when written with indices as described in this
paper, lead to no degrees of freedom for the little group and one may
suspect they do not correspond to propagating degrees of freedom. One
word of caution is required in relation to  this last statement as it
depends  on writing  all generators with $3l$ indices, however,   one can
change the number of indices with the epsilon symbol and only the
dynamics can determine which fields actually propagate. Only once one has
the dynamics can one determined what the actual propagating degrees of
freedom are.
\par
The corresponding generators
in the IIB theory follow a similar pattern and interpreta\-tion and it is
likely that the results hold for any non-linear realisation based on any
very extend algebra $G^{+++}$.
\par
Given the results in this, and previous papers on
$E_{11}$, there is now considerable evidence that
$E_{11}$ is closely associated with the maximal supergravity theories.
Indeed,  it would  seem increasingly likely that the original conjecture
of reference [1] that $E_{11}$ is a symmetry of a suitably decorated 
formulation of  eleven dimensional supergravity is true. Since the
maximal ten dimensional supergravity theories are the complete low energy
effective actions for the corresponding superstrings this result also
encourages the belief that the underlying theory has an $E_{11}$
symmetry, albeit of a discrete kind.
\par
The dual generators are only a small part of the infinite number of
generators of the adjoint representation of $E_{11}$. However, as
explained in the introduction a number of the non-dual   generators,
although non-propagating,  have a physical meaning. For example, the
level  four generator
$R^{a,b,c_1\ldots c_{10}}$ of the  eleven dimensional theory allows for
the emergence  of the massive IIA supergravity theory upon reduction to
ten dimensions. One can expect that this is a general phenomenon, namely
the gauged, sometimes called massive,  supergravities that occur in lower
dimensions can be thought of as arising from fields   beyond the
supergravity approximation in the eleven dimensional $E_{11}$ non-linear
realisation. Indeed, it would be interesting to see if a precise
correspondence can be made between such fields and the gauged
supergravities in say three dimensions. Thus one expects to find an
 interpretation for many of the non-dual generators and these
results encourage the belief that all the fields in the non-linear
realisation of $E_{11}$  have physical meaning.
\par
In this paper we have only shown that the field content of the non-linear
realisations corresponds  to all possible dual fields. However, it would
be good to show that the dynamics of the non-linear realisation actually
leads to a formulation with all possible  dual field incorporated in the
correct way. It is possible that  the dual generators and the generator
with $u=0$ of equation (18)
form a closed subalgebra which is a  truncation of the
$E_{11}$ with the generators of equation (11) added. If this were the
case there would be  a manifest description of the duality symmetries of
the on-shell degrees of freedom without the non-dual fields.
\medskip
{\bf Acknowledgments}
Peter West would like to thank Rafael Montemayor for important discussions
at an early stage of this work and hospitality at the centre for physics
at Bariloche in January 2005 and at the physics department of the
University of Canterbury at  Christchurch, New Zealand in November 2006.
He would   also like to thank  Andrew Pressley for discussions on group
theory.   Fabio Riccioni would like to thank E. Bergshoeff for
discussions. The  research of Peter West was supported by a PPARC senior
fellowship PPA/Y/S/2002/001/44. The work of both authors is also
supported by   a  PPARC rolling grant   PP/C5071745/1 and  the EU Marie
Curie, research training network grant HPRN-CT-2000-00122.
\medskip
{\bf References}
\medskip
\item{[1]} P. West, {\sl $E_{11}$ and M Theory}, Class. Quant.
Grav. {\bf 18 } (2001) 4443,  hep-th/010408.
\item{[2]} P.~C. West, {\sl Hidden superconformal symmetry in {M}
   theory },  JHEP {\bf 08} (2000) 007, {\tt hep-th/0005270}
\item{[3]} I. Schnakenburg and P. West, {\sl Kac-Moody Symmetries of
IIB supergravity}, Phys. Lett. {\bf B 517} (2001) 137-145, {\tt
hep-th/0107181}
\item{[4]} A. Kleinschmidt, I. Schnakenburg and P. West, {\sl
Very-extended Kac-Moody algebras and their interpretation at low
levels}, {\tt hep-th/0309198}
\item{[5]} P. West, {\it  The  IIA, IIB and eleven dimensional
theories and their common $E_{11}$ origin}, hep-th/0402140.
\item{[6]} J. Schwarz and  P. West {\it Symmetries and Transformations of
    chiral N=2, D=10 super\-gravity}, Phys. Lett. {\bf B126} (1983), 301;
    J. Schwarz, {\it Covariant field equations of chiral N=2 D=10
    supergravity}, Nucl. Phys. {\bf B226} (1983), 269; P. Howe and  P.
West,    {\it The complete N=2, d=10 supergravity}, Nucl. Phys. {\bf B238}
    (1984) 181.
\item{[7]} E. Bergshoeff, Mess de Roo, S. Kerstan and F. Riccioni, {\it
IIB Supergravity Revisited}, hep-th/0506013.
\item{[8]} P. West, {\it  $E_{11}$, ten forms and supergravity }, JHEP
0603 (2006) 072, hep-th/0511153.
\item{[9]}  P. Meessen and T. Ortin, {\it An SL(2,Z) multiplet of nine
-dimensional type II super\-gravity theories}, Nucl. Phys. B {\bf 541}
(1999) 195, hep-th/9806120;
G. Dall'Agata, K. Lechner and M. Tonin, {\it D=10, N=IIB supergravity;
Lorentz-invariant actions and duality}, JHEP {\bf 9807} (1998) 017,
hep-th/9806140;
\item{[10]}  A. Sagnotti, {\it Open strings and their symmetry groups},
hep-th/0208020.
\item{[11]} Eric A. Bergshoeff, Mees de Roo, Sven F.
Kerstan, T. Ortin,  Fabio Riccioni, {\it  IIA Ten-forms and the Gauge
Algebras of Maximal Supergravity Theories},   JHEP 0607 (2006) 018,
hep-th/0602280.

\item{[12]} M. R. Gaberdiel, D. I. Olive and P. West, {\sl A
class of Lorentzian Kac-Moody algebras}, Nucl. Phys. {\bf B 645}
(2002) 403-437,  hep-th/0205068.
\item{[13]}  T. Damour, M. Henneaux and H. Nicolai, {\sl $E_{10}$ and a
``small tension expansion'' of M-theory}, Phys. Rev. Lett. {\bf 89}
(2002) 221601, {\tt hep-th/0207267}
\item{[14]} P. West, {\it Very Extended $E_8$ and $A_8$ at low levels,
Gravity and Supergravity}, Class. Quant. Grav. 20 (2003) 2393-2406,
hep-th/0212291.
\item{[15]}  V. Kac, {\it ``Infinite Dimensional Lie
Algebras"}, Birkhauser, 1983.
\item{[16]}  T. Fischbacher and H.  Nicolai, {\it
Low Level Representations for E10 and E11},
Contribu\-tion to the Proceedings of the Ramanujan International
Symposium on Kac-Moody Algebras and Applications, ISKMAA-2002, Jan.
28-31, Chennai, India. hep-th/0301017.
\item{[17]} N.D. Lambert, P.C. West, {\it Coset Symmetries in
Dimensionally Reduced Bosonic String Theory},  Nucl.Phys. B615 (2001)
117-132, hep-th/0107209.
\item{[18]} F. Englert, L. Houart, A. Taormina, P. West,

{\it  The Symmetry of M-Theories},  JHEP 0309 (2003) 020,
hep-th/0304206.

\item{[19]} A. Kleinschmidt and P. West, {\it  Representations of G+++ and
the role of space-time},  JHEP 0402 (2004) 033,  hep-th/0312247.

\end